\documentstyle[12pt,aasms,flushrt]{article}

\def\aa#1{AA, #1}

\def\aj#1{AJ, #1}
\def\apj#1{ApJ, #1}
\def\apjl#1{Ap. J. (Letters), #1}
\def\apjsup#1{ApJS, #1}

\def\mn#1{MNRAS, #1}

\def\ten#1{\ifmmode 10^{#1}\else $10^{#1}$\fi}
\def\gsim{\buildrel > \over \sim}

\def\approxgt{\lower.2em\hbox{$\buildrel > \over \sim$}}
\def\approxlt{\lower.2em\hbox{$\buildrel < \over \sim$}}
\def\uline#1{$\underline{\hbox{\kern #1in}}$ }
\def\sqr#1#2{{\vcenter{\hrule height .#2pt
          \hbox{\vrule width .#2pt height#1pt \kern#1pt
                  \vrule width .#2pt}
          \hrule height .#2pt}}}

\def\sc{^{\rm s}}

\def\secpoint{''\mskip-7.6mu.\,}

\def\88in{$88^{\prime\prime}$}
\def\24in{$24^{\prime\prime}$}

\def\ang{~{\rm \AA}}

\def\kms{~{\rm km\ sec}^{-1}}
\def\cm2{~\rm cm^{-2}}

\def\chisq{\ifmmode {\chi^2}\else {$\chi^2$}\fi}

\def\lal{{\rm Ly}\alpha}
\def\lbet{{\rm Ly}\beta}

\def\qno{q_0}

\def\hno{{\rm H}_0}

\def\ten#1{\times 10^{#1}}

\def\carbon4{{\ion{C}{4}}}
\def\carbon2{{\ion{C}{2}}}
\def\c2{{\ion{C}{2}}}
\def\si4{{\ion{Si}{4}}}
\def\h1{{\ion{H}{1}}}

\begin{document}

\title{Metal Enrichment and Ionization Balance in the Lyman $\alpha$\ Forest\\at $z = 3$.}
\author{Antoinette Songaila\altaffilmark{1} \& Lennox L. Cowie\altaffilmark{1}} 
\affil{Institute for Astronomy, University of Hawaii, 2680 Woodlawn Dr.,
  Honolulu, HI 96822\\}

\altaffiltext{1}{The authors were visiting astronomers at the W.  M. Keck
Observatory, jointly operated by the California Institute of Technology and
the University of California.  The research was supported at the University of
Hawaii by the State of Hawaii.}

\vskip 1in

\centerline{Accepted for publication in {\it Astronomical Journal\/}, May 14, 1996 }

\begin{abstract}

The recent discovery of carbon in close to half of the low neutral hydrogen
column density [$N({\rm H~I}) > 3\ten{14}\cm2$] Lyman forest clouds toward $z
\sim 3$\ quasars has challenged the widely held view of this forest as a
chemically pristine population uniformly distributed in the intergalactic
medium, but has not eliminated the possibility that a primordial population
might be present as well.  Using extremely high signal-to-noise observations
of a sample of quasars  we now show that 
\ion{C}{4} can be found in 75\% of clouds with $N({\rm H~I}) > 3\ten{14}\cm2$\
and more than 90\% of those with $N({\rm H~I}) > 1.6\ten{15}\cm2$.  Clouds
with $N({\rm H~I}) > 10^{15}\cm2$\ show a narrow range of ionization ratios,
spanning less than an order of magnitude in \ion{C}{4}/\ion{H}{1},
\ion{C}{2}/\ion{C}{4}, \ion{Si}{4}/\ion{C}{4} and \ion{N}{5}/\ion{C}{4}, and
their line widths require that they be photoionized rather than collisionally
ionized. This in turn implies that the systems have a spread of less than an
order of magnitude in both volume density and metallicity.  Carbon is seen to
have a typical abundance of very approximately $10^{-2}$\ of solar and Si/C
about three times solar, so that the chemical abundances of these clouds are
very similar to those of Galactic halo stars.  \ion{Si}{4}/\ion{C}{4}
decreases rapidly with redshift from high values ($> 0.1$) at $z > 3.1$,
a circumstance which we interpret as a change in the ionizing spectrum as the
intergalactic medium becomes optically thin to He$^+$\ ionizing photons.  Weak
clustering is seen in the \ion{C}{4} systems for $\Delta v < 250\kms$,
which we argue provides an upper limit to the clustering of \ion{H}{1}
clouds.  If the clouds are associated with galaxies, this requires a 
rapid evolution in galaxy clustering between $z = 3$\ and $z = 0$.

\end{abstract}

\section{Introduction}

A clear theoretical paradigm is beginning to emerge in which the forest of
$\lal$\ absorption lines seen in quasar spectra is interpreted in terms of the
development of structure in the intergalactic gas (Cen et al.\ 1994).  This
model subsumes many aspects of previous models (Rees 1995) but places the
observations in the much broader intellectual context of the formation of
galaxies and large scale structure and of the evolution of the intergalactic
medium as a whole.

This type of model has considerable predictive power and it is already clear
that it can explain in broad terms the hydrogen properties of the forest
clouds --- that is, the column density distribution, $b$-values and turbulence
of the $\lal$\ forest lines (Zhang, Anninos \& Norman 1996) but, given the
uncertainties in cosmology, the perturbation spectrum and the evolution of the
ionizing field, the coarse information available in the \ion{H}{1} data may
not sufficiently challenge the theory.  However, the models also make detailed
predictions of the ionization balance (collisional and photoionization) in
heavier elements (Haehnelt et al.\ 1996) that are now testable with the
detection of metals in many forest clouds (Cowie et al.\ 1995 [Paper~I],
Tytler et al.\ 1995).

The presence of metals is also of considerable interest in itself.  If forest
clouds are associated with galaxies, large radii ($\sim 200 h^{-1}~{\rm kpc}$)
are required to account for the number density of $N({\rm H~I}) >
10^{15}\cm2$\ clouds at $z = 3$, and it appears unlikely that star formation
in the galaxy itself could contaminate this entire region.  The remaining
possibilities would seem to be that the metals are being produced by {\it in
situ\/} star formation in the clouds themselves, perhaps as an early stage in
the galaxy formation process, or that they are formed in an even earlier stage
of metal production at much higher redshift that has uniformly enhanced the
intergalactic medium.  Since this latter (Population~III) process presumably
would occur on a sub-galactic scale, we might expect it to produce a more
uniform metallicity at the scale of the clouds than {\it in situ\/} star
formation would produce; in contrast, we would expect {\it in situ\/}
enrichment to depend heavily on the properties of individual clouds and
perhaps to be strongest in higher density clouds.

We shall show in this paper that $N({\rm C~IV}) \ge 10^{12}\cm2$\ in 90\% of
clouds with $N({\rm H~I}) > 1.6\ten{15}\cm2$\ and $N({\rm C~IV}) \ge 5\times
10^{11}\cm2$\ in 75\% of clouds with $N({\rm H~I}) > 3.0\ten{14}\cm2$.  The
fraction of detections is consistent with the hypothesis that all clouds have
a uniform distribution of \ion{C}{4}/\ion{H}{1} as a function of $N({\rm
H~I})$, but still leaves open the possibility that as much as 50\% of clouds at
these column densities may be primordial.  The heavy element properties of the
$N({\rm H~I}) > 10^{15}\cm2$\ clouds are remarkably uniform and may pose a
significant challenge to the structure development models which predict a
wide range of densities and temperatures in the clouds.  This uniformity also
favors the pre-enrichment hypothesis for the origin of the metals.

The forest metal lines also provide excellent probes of the intergalactic
ionizing flux and its evolution.  Since they are generally weak lines, their
column densities can be measured accurately by profile fitting and there are
no considerations of neutral hydrogen optical depth as there are for higher
column density clouds.  We find that there is a significant evolution in
\ion{SI}{4}/\ion{C}{4} as a function of redshift, with an abrupt decrease in
this quantity at $z < 3.1$.  The data can be understood in terms of a
photoionization model in which the He$^+$\ edge becomes optically thin below
$z = 3.1$.  The one cloud close to a quasar ($\Delta v < 1000\kms$) is found
to have significantly higher ionization parameters but also to have anomalous
metallicity, in agreement with the findings of Petitjean, Rauch \& Carswell
(1994) for higher column density systems.

Finally, we consider the velocity clustering of the \ion{C}{4} systems (e.g.
Fernandez-de Soto et al.\ 1996).  The clustering extends to $\Delta v =
250\kms$; if forest clouds are assumed to trace galaxy clustering at these
redshifts, this requires a rapid evolution in galaxy clustering from $z = 3$\
to $z = 0$.

\section{Observations}

We draw primarily on two extemely high sensitivity, high resolution ($R =
36,000$) spectra of the $z = 3.4$\ quasar Q0014+813, and of the $z = 3.6$\
quasar Q1422+231, that we have obtained with the HIRES spectrograph on the
Keck 10~m telescope.  Total integration times were 7.75~hours on Q0014+813 and
8.3~hours on Q1422+231.  In each case the spectrograph was used in three
separate configurations to provide complete wavelength coverage of all
\ion{C}{4} absorption longward of the Lyman forest as well as the
corrresponding $\lal$\ and $\lbet$\ portions of the spectra.  Q1422+231 is
gravitationally lensed but the major components all lie within the $1\secpoint
1$\ slit used in the observations; the combined source is therefore extremely
luminous ($g = 15.8$; Yee \& Ellingson 1994) and provided a truly exceptional
spectrum, illustrated in Figure~1, with S/N = 160 per resolution element in
the segment from $6300\ang$\ to $6400\ang$\ shown in the upper panel.  The
spectrum of Q0014+813 has poorer S/N by about, on average, a factor of two.
For certain of the measurements, where a larger sample was needed, we included
spectra of a larger collection (Songaila 1996) of $z \sim 3$\ quasars; this is
summarized in the legend of Fig.~4.  Details of the reduction process are
given in Paper~I and Hu et al.\ (1995), with a more extensive discussion in
Songaila (1996).

\section{Fractions of C~IV Detections in Forest Clouds}

\subsection{Introduction}

In Paper~I and Tytler et al.\ (1995) it was shown that roughly 50\% of
$\lal$\ forest clouds with $N(\rm H~I) \gsim 3\ten{14}\cm2$\ have
\ion{C}{4} absorption with $N(\rm C~IV) > 10^{12}\cm2$.  A critical
question is whether the clouds that are undetected in \ion{C}{4} are simply
below the detection limit or whether at these \h1 column densities there are
two classes of forest cloud --- one that contains metals and another that is
chemically pristine.  In order to try to understand this issue we have looked
at two further samples.  In section~3.2 we consider a higher-$N(\rm H~I)$\
selection [$N(\rm H~I) \ge 1.6\ten{15}\cm2$] and in section~3.3 we use the
very sensitive Q1422+231 observations to determine the \ion{C}{4} fraction in
a $\lal$-selected sample (i.e. $N(\rm H~I) \gsim 3\ten{14}\cm2$).

\subsection{Lyman $\beta$\ selection}

For Q1422+231 we first selected all Lyman forest cloud components for
which $r_{\nu} \equiv \exp (-\tau_{\nu}) \le 0.02$.  Here $r_{\nu}$\ is the
residual flux at the base of the line and $\tau_{\nu}$\ is the
corresponding optical depth.  The search detected 62 clouds in a redshift
interval from $z = 2.65$, where \ion{C}{4} lies redward of the $\lal$\ forest,
to the quasar's redshift $z = 3.625$\ (Fig.~1).  This constitutes the
$\lal$\ sample we discuss in the next subsection.  Fig.~2 is an atlas of
these lines along with the corresponding \ion{C}{4} ($\lambda 1548,
1550\ang$) lines and $\lbet$\ (redward of $z = 2.92$; see below).

For clouds with $z > 2.92$\ the Lyman~$\beta$\ line lies sufficiently redward
of the Lyman edge, at $4000\ang$, corresponding to the $z = 3.38$\ Lyman limit
system, that the signal-to-noise at the $\lbet$\ position is adequate for
applying our selection criteria, and between here and the quasar redshift we have
drawn a subsample of clouds that are saturated in both $\lal$\ and $\lbet$.
This method of selection is illustrated in Fig.~3.  The selection chooses
clouds that have a higher column density than a $\lal$\ selection would
choose, in roughly the ratio (= 5.2) of the $\lal$\ to $\lbet$\ oscillator
strengths; this picks out clouds with $N(\rm H~I)
\gsim 1.6\ten{15}\cm2$\ for $b$-values typical of forest clouds.

In order to supplement this sample we applied the same procedure to the
spectra from six other quasars: these are the four quasars of Paper~I plus
Q1159+123 and Q2126$-$158, which were observed subsequently.  Except for
Q0014+813, these spectra do not extend to the quasars' \ion{C}{4} emission,
and only a narrower redshift range could be used in each case; these redshift
ranges are summarised in the legend of Fig.~4.

For each cloud selected, as long as the \ion{C}{4} lay redward of the $\lal$\
forest, the column density of \ion{C}{4} was measured by profile fitting
to the \ion{C}{4} $1548\ang$\ and $1550\ang$\ lines.  All the lines are
unsaturated and there is good agreement between the two members of the
doublet.  The sensitivity limit for \ion{C}{4} is slightly variable from
quasar to quasar, but for $N(\rm C~IV) > 10^{12}\cm2$, both members of the
doublet can generally be detected at the $2\sigma$\ level for $b =
10\kms$, and we adopt this as our sensitivity limit.  

There are 41 clouds in the sample of $\lbet$-saturated clouds, of which 36
(88\%) are detected in
\ion{C}{4} .  This is shown in Fig.~4 where we plot $N(\rm C~IV)$\ versus
redshift; the undetected systems are shown at $N(\rm C~IV) = 10^{12}\cm2$.
There is no significant evolution of the $N(\rm C~IV)$\ distribution as a
function of redshift.  Some fraction (roughly half) of the 5 systems that are
not detected will correspond to false systems, where there is $\lal$\
contamination of the $\lbet$\ position, so that the actual detection fraction
lies between 90\% and 100\%.  This is our first conclusion: essentially all
$\lal$\ forest clouds with $N(\rm H~I) \gsim 1.6\ten{15}\cm2$\ contain metals.

The median column density of the \ion{C}{4} lines is $4.3\ten{12}\cm2$,
and the mean column density is $1.61\ten{13}\cm2$\ if the five undetected
systems are included at $N(\rm C~IV) = 10^{12}\cm2$, and is
$1.59\ten{13}\cm2$\ if these five are assigned zero column density.  If
we assume that the hydrogen column density distribution is of the form
$n(N(\rm H~I))dN \sim N^{-1-\beta}dN$, then, for $\beta = 0.7$, the median
$N(\rm H~I)$\ is $2.6\ten{15}\cm2$\ and the mean value $9.7\ten{15}\cm2$,
where we have assumed an upper cutoff value of $10^{17}\cm2$\ in
calculating the latter.  The ratio of the medians gives
\ion{C}{4}/\ion{H}{1} $= 1.6\ten{-3}$; the ratio of the means 
gives an identical result.

Figs.~4(b) and (c) show the corresponding diagrams for the \si4
($\lambda\lambda 1393, 1402\ang$) and \ion{N}{5} ($\lambda\lambda 1238,
1242\ang$) doublets.  Because of the short wavelength of \ion{N}{5} there is
only a very small number of systems that lie redward of the $\lal$\ forest.
Fig.~4(d) shows the corresponding diagram for
\ion{C}{2} ($\lambda 1334.5\ang$); since there is no corroborating additional
line in this case a small fraction of the detections may be spurious.  38\% of
the clouds are detected in \si4 and the mean column density lies in the range
$(1.5 - 1.8)\ten{12}\cm2$, depending on whether undetected clouds are given a
value of 0 or $5\ten{11}\cm2$.  The corresponding values for \ion{N}{5} are
17\% and $(0.2 - 1.1)\ten{12}\cm2$, and for
\ion{C}{2} are 24\% and $(1.7 - 2.5)\ten{12}\cm2$.  The ratios of these
values to that of \ion{C}{4} in the mean are \si4/\ion{C}{4} =
[0.09,0.11], \ion{N}{5}/\ion{C}{4} = [0.01,0.07] and
\ion{C}{2}/\ion{C}{4} = [0.11,0.16].

In contrast to \ion{C}{4}, both the \si4 and \ion{C}{2} appear to show a
trend to increasing values at higher $z$.  This has been noted previously
in studies of equivalent widths of the higher column density metal lines
(e.g.\ Bergeron \& Ikeuchi 1990).  We will return to this point in the next
section.

\subsection{Lyman $\alpha$\ selection}

The extremely high S/N of the Q1422+231 observations has made it possible to
search for \ion{C}{4} absorption in the entire $\lal$\ selected sample of
Fig.~2 to a considerably lower column density than was possible for the
quasars of Paper~I.  We show $N(\rm C~IV)$\ versus redshift for all the
$\lal$\ selected clouds in Q1422+231 in Fig.~5, where we have adopted a
conservative limit of $N(\rm C~IV) = 5\ten{11}\cm2$\ at which both components
of the doublet will be detected above the $2\sigma$\ level.

As with the $\lbet$-selected sample, there is no obvious trend with
redshift.  $(75 \pm 10)$\% of the clouds (50 out of 75) are detected in
\ion{C}{4} with a median $N(\rm C~IV) = 1.4\ten{12}\cm2$\ and a mean of
$[9.5,9.6]\ten{12}\cm2$.  Assuming a median $N(\rm H~I)$\ of $5\ten{14}\cm2$\
and a mean of $4\ten{15}\cm2$\ we obtain a median ${\rm C~IV/H~I} =
2.8\ten{-3}$\ and a mean of ${\rm C~IV/H~I} = [2.4,2.4]\ten{-3}$, similar to
the values obtained in Paper~I and very slightly higher than those
derived from the $\lbet$-selected sample.

If the $N(\rm C~IV)$\ threshold is raised to $10^{12}\cm2$\ we detect 41 out
of 67 clouds, or $(61 \pm 9)$\%, compared to 17 out of 33, or $(52 \pm
13)$\%, in the Paper~I sample.  Combining the two data sets we find that
58 out of 100 clouds, or $(58 \pm 8)$\%, are detected above $N(\rm C~IV) =
10^{12}\cm2$. 

\subsection{Detection fractions}

In Fig.~6 we show the fraction of clouds detected in \ion{C}{4} as a
function of the variable $x \equiv N({\rm C~IV)_{det}} /N(\rm
H~I)_{select}$\ for the $\lal$-selected sample, where $N(\rm
H~I)_{select} = 3\ten{14}\cm2$.  Here $N(\rm C~IV)_{det}$\ is our
detection limit for \ion{C}{4}, which we can arbitrarily adjust above our
observational limit, and $N(\rm H~I)_{select}$\ is our
selection limit for \ion{H}{1}.  For $N(\rm C~IV)_{det} \ge 10^{12}\cm2$\
we have used the data of Q1422+231 augmented by the Paper~I data, whereas
the $N(\rm C~IV) < 10^{12}\cm2$\ data point is from Q1442+231 alone.

If the column density distribution of \ion{H}{1}, $n(N(\rm H~I))$, can be
described approximately by a power law over the range $3\ten{14}\cm2 \le
N(\rm H~I) \le 10^{17}\cm2$, and the distribution function of
\ion{C}{4}/\ion{H}{1} is invariant as a function of $N(\rm H~I)$, then the
detection fraction is a function only of the single dimensionless
variable, $x$.  In this case the detection fractions as a function of
$x$\ should be identical for the $\lal$\ and $\lbet$\ samples.  Fig.~6
illustrates this agreement (the solid line is the $\lbet$\ sample),
showing that the data is completely consistent with the hypothesis that
all clouds are metal-enriched with a distribution function in
\ion{C}{4}/\ion{H}{1} which is invariant to $N({\rm H~I})$, at least around 
these values of $N(\rm H~I)$.

We can also use the data to estimate an upper limit to the fraction of clouds
between $N(\rm H~I) = 3\ten{14}\cm2$\ and $1.6\times10^{15}\cm2$\ that can be
primordial.  If the \ion{H}{1} distribution is a power law with $\beta = 1.7$,
then 69\% of the clouds in an $N(\rm H~I) > 3\ten{14}\cm2$\ sample will lie in
the range $3\ten{14} - 1.6\ten{15}\cm2$.  (In fact the fraction of lower
column density clouds could be higher at these \ion{H}{1} column densities
because of the more complex shape of $n(N(\rm H~I))$\ [Petitjean et al.\ 1993,
Hu et al.\ 1995].)  Of the $(75 \pm 10)$\% of the $\lal$-selected clouds
detected in \ion{C}{4}, 31\% will be contained in the $\lbet$\ sample in which
essentially all clouds are detected in
\ion{C}{4}.  Of the remaining 69\%, at least $(44 \pm 10)$\% are detected
in \ion{C}{4}, so that a minimum of $(63 \pm 14)$\% of the clouds in this
column density range are detected in \ion{C}{4}.  

It can be seen from Fig.~6 that if the clouds are all metal-enriched, a
complete identification of the $\lal$\ sample would require a sensitivity
limit corresponding to $x = 5\ten{-4}$\ or $N(\rm C~IV) \sim 10^{11}\cm2$\
which would be hard to achieve even for a quasar as luminous as Q1422+231.
Similarly, even a 50\% identification of an $N(\rm H~I) \ge 10^{14}\cm2$\
sample would require a sensitivity of $N(\rm C~IV) = 4\ten{11}\cm2$.  This
means that it will remain very difficult to determine unambiguously the
fraction of metal-enriched clouds at column densities much below $10^{15}\cm2$, and
it remains at least conceptually possible that a large fraction of very low
$N(\rm H~I)$\ clouds could be metal-poor or pristine.

\section{Ionization Balance}

\subsection{Introduction}

As we showed in the previous section, the $\lal$\ forest clouds in the
$N({\rm H~I}) \ge 1.6\ten{15}\cm2$\ sample are most easily detected in
\ion{C}{4} and less so in \ion{C}{2}, \ion{Si}{4}, and \ion{N}{5}.  In
the present section we investigate the ionization ratios in individual clouds
with $N(\rm H~I) > 5\ten{14}\cm2$.  These turn out to have a relatively small
spread, which suggests two possibilites.  One is that these clouds are highly
invariant in both their average metallicity and in their internal properties,
in particular the suitably weighted average internal density.  Alternatively,
they could be very strongly selected from a more heterogeneous population.  In
section~4.2 we consider the distribution of the average \ion{C}{4}/\ion{H}{1}
in $N(\rm H~I) \ge 5\ten{14}\cm2$\ forest clouds, and in section~4.3 we
investigate the internal spread of \ion{C}{4}/\ion{H}{1} in the velocity
components of individual clouds.  In section~4.4 we look at the ratios
\ion{Si}{4}/\ion{C}{4},
\ion{C}{2}/\ion{C}{4} and
\ion{N}{5}/\ion{C}{4} and how they restrict the possibility of
photoionization models, and in section~4.5 we look
at the ionization balance based on a larger list of species in several of
the strongest clouds.  Finally, in section~4.6 we investigate the
evolution of the ion ratios, showing that there is a strong increase of
\ion{Si}{4}/\ion{C}{4} with redshift, which has important implications for
understanding the evolution of the He$^+$\ opacity in the IGM.

\subsection{C~IV/H~I}

At sufficiently high redshift we can measure the properties of the
$\lal$-saturated clouds by tracing down through the Lyman series.  We have
measured the \ion{H}{1} properties of all clouds above $z = 3.135$\ in
Q1422+231 and all clouds toward Q0014+813 above $z = 2.95$.  From these we
have chosen all clouds with $N(\rm H~I) > 5\ten{14}\cm2$\ in Q1422+231 and
$N(\rm H~I) > 10^{15}\cm2$\ in Q0014+813 where the \ion{C}{4} sensitivity is
about a factor of two poorer.  The properties of these clouds are summarised
in Tables~1 and 2.  We also summarise in Table~3 the average properties of all
the partial Lyman limit systems (PLLSs)in the observed quasars (Songaila
1996).  In Fig.~7 we display \ion{C}{4}/\ion{H}{1}, averaged over all the
velocity components in an individual cloud, for all the objects in Tables~1, 2
and 3.  All 8 of the PLLSs are detected.  At $10^{17} < N(\rm H~I) <
10^{18}\cm2$, the median \ion{C}{4}/\ion{H}{1} is $1.3\ten{-4}$\ and the mean
is $2.3\ten{-4}$.  For $N(\rm H~I)$\ in the range $10^{15} - 10^{17}\cm2$, 14
out of 18 clouds are detected.  The mean \ion{C}{4}/\ion{H}{1} is
$3.5\ten{-3}$\ and the median value is $3\ten{-3}$.  At $N(\rm H~I) =
5\ten{14} - 10^{15}\cm2$, 5 out of 9 clouds are detected, with a median
\ion{C}{4}/\ion{H}{1} of $2\ten{-3}$.  The values for the forest clouds are
quite similar to those derived in Section~2 but are now free of assumptions
about the distribution of \ion{H}{1} column densities, at the cost of having a
rather reduced sample.

If we assume for the moment that each cloud is homogeneous and
photoionized by a metagalactic flux of specified shape, then
\ion{C}{4}/\ion{H}{1} is a function only of the ionization
parameter, $\Gamma$\ (the ratio of the number density of hydrogen ionizing
photons to the electron density) and the metallicity in the cloud.  In Fig.~7
we show curves of \ion{C}{4}/\ion{H}{1} versus $N({\rm H~I})$\ from Bergeron
\& Stasinska (1986) for $\Gamma = 10^{-2.7}$, $10^{-1.7}$, and $10^{-0.7}$\ 
and a carbon abundance of $10^{-2}$\ of solar.  For $\Gamma = 10^{-2.5} -
10^{-1.5}$\ (section 4.4), which provides the best match to the ionization of
other species, the carbon abundances are approximately $10^{-2}$\ of solar,
though there could be a spread of at least an order of magnitude in individual
clouds.  It is extremely hard to deviate much from this abundance, and in
particular to reduce the carbon abundance, by changing such assumptions as the
internal homogeneity of the clouds because \ion{C}{4}/\ion{H}{1} is quite
insensitive to $\Gamma$\ over this range, and close to its peak value.  We
will consider this further in section~4.4.

\subsection{C~IV/H~I in individual velocity components}

The \ion{C}{4}/\ion{H}{1} values of the previous subsection are 
averaged over the sometimes complex velocity component structure of an
absorption line system, and it is important to understand if the small spread
in \ion{C}{4}/\ion{H}{1} is an artefact of this averaging or if it continues
to hold in the individual velocity components of a system.  To look further
at this we have determined hydrogen column densities in individual velocity
components of the systems in Q1422+231 by fitting the high order Lyman lines,
where possible, and have compared these with \ion{C}{4}.  
A detailed description of the individual systems in Q1422+231 may be found in
Appendix~A.  

The stronger \ion{H}{1} components in each system contain \ion{C}{4}
with \ion{C}{4}/\ion{H}{1} between $5\ten{-4}$\ and $1.5\ten{-2}$, the
majority lying near $3\ten{-3}$\ (Fig.~8a).  The bulk of the \ion{C}{4}
occurs in these components, with the result that these \ion{C}{4}/\ion{H}{1}
ratios are very similar to the average values of the previous subsection.
Surrounding them at relative velocities as high as $350\kms$\ (Fig.~8b) are
components with much higher values of
\ion{C}{4}/\ion{H}{1}, in the range $10^{-2}$\ to $10^{-1}$, the majority
being near the latter value.

The presence of very narrow \ion{C}{4} (as in the $z = 3.513$\ system) and,
more generally, a comparison of the $b$-values of \ion{C}{4} and those of
\ion{H}{1}, suggests that much of the broadening even in the strongest
\ion{H}{1} components is turbulent rather than thermal.  This in turn implies
that measured \ion{H}{1} $b$-values of 20 -- $28\kms$\ provide a strong upper
bound of 50,000~K to the temperature in strong
\ion{H}{1} components.  This upper limit also applies to most of the
higher velocity components.  At these temperatures, if the ionization balance
were governed by collisional ionization, \ion{C}{2} would exceed
\ion{C}{4} even in a gas cooling from high temperature (e.g.\ Shapiro \&
Moore 1976).  It would therefore appear that the gas must be predominantly
photoionized.  The more highly ionized high velocity components must then be
one to two orders of magnitude lower in volume density than the higher
column density components.

\subsection{Ionization balance}

In Figs.~9a through 9c we show values of \ion{Si}{4}/\ion{C}{4},
\ion{C}{2}/\ion{C}{4} and \ion{N}{5}/\ion{C}{4} versus $N({\rm C~IV})$\
for all measurable systems in the quasar sample for which $N({\rm H~I}) >
5\ten{14}\cm2$\ and the lines of both species lie longward of the Lyman
forest.  We have also shown (open symbols) the values in the various
partial Lyman limit systems.

All the systems (forest and PLLS) are consistent with having
\ion{Si}{4}/\ion{C}{4} in the range 0.03 to 0.4.  (As we shall discuss in
section~4.6, the spread at a given redshift is smaller.)
\ion{C}{2}/\ion{C}{4} lies between 0.02 and 0.3, except for the strong
PLLS system in Q0636+680 for which \ion{C}{2}/\ion{C}{4} = 2.  Finally,
the small number of \ion{N}{5}/\ion{C}{4} measurements yield two values of
0.02 and 0.05, and two comparable upper limits.  These values are
again consistent with the discussion of the broader samples of section~3.
Distinguishing by velocity component again gives similar ionization ratios in
the strongest \ion{H}{1} components.

If we assume that the highest column density component of the optically thin
forest clouds can be parameterised by a single value of the ionization
parameter, $\Gamma$, and that the ionizing spectrum has a specified form, we
can determine the ionization parameter from these ion ratios (e.g. Bergeron \&
Stasinska 1986, Chaffee et al.\ 1986, Steidel 1990, Donahue \& Shull 1991).
The
\ion{C}{2}/\ion{C}{4} value is most useful for this since it is
independent of any assumption about abundance and is insensitive to the
exact shape of the high-energy end of the ionizing spectrum and, in
particular, to whether or not there is a substantial break across the
He$^+$\ edge at 54~eV.  Using the CLOUDY code (Ferland 1993) we find that
$\Gamma$\ must be in the range $10^{-2.4}$\ to $10^{-1.5}$\ to produce the
observed \ion{C}{2}/\ion{C}{4} values, for a $\nu^{-1.5}$\
power law;  these values are similar to those inferred from higher
column density systems in the work referred to above.  This last result can
also be seen by comparing the values of \ion{C}{2}/\ion{C}{4} in the PLLSs
with those in the lower column density systems (Fig.~9b).

The high ion ratios are sensitive to the choice of the shape of the
high-energy end of the ionizing spectrum; unfortunately this is rendered quite
uncertain by the effect of the He$^+$\ opacity in the IGM and the poorly known
relative contributions of galaxies and AGN.  We follow Giallongo \& Petitjean
(1994) by characterising the spectrum as a broken power law, retaining the
$\nu^{-1.5}$\ slope but dropping the normalization by a factor $B$\ at the
He$^+$\ edge.  We have then computed a suite of models with the CLOUDY code
corresponding to values of $B$\ from 1 to $\infty$.

We show the resultant \ion{Si}{4}/\ion{C}{4} ratios versus
\ion{C}{2}/\ion{C}{4} in Fig.~10, with small symbols denoting $z <3.1$\
and large symbols corresponding to $z > 3.1$.  Intriguingly, these
models do not provide a good fit for Si/C in a solar abundance ratio
(left panel of Fig.~10) but, as with the low metallicity stars in our
Galaxy (e.g.\ Timmes et al.\ 1996), require an enhancement of 3 in the
$\alpha$-process silicon over the Fe-process carbon (right panel);  this
suggests a weighting to higher masses in the metal-producing stars.
This is the second conclusion of the paper:  Si/C ratios in the forest
clouds are similar to those in low metallicity stars in our Galaxy ---
i.e. about a factor of 3 higher than solar.

The high \ion{Si}{4}/\ion{C}{4} values seen in $z > 3.1$\ clouds even
in systems with small \ion{C}{2}/\ion{C}{4} (that is, high ionization)
require that the strength of the break be very large in these clouds
($B \ge 100$) to inhibit silicon from moving to higher ionization levels.
This in turn means that there are very few He ionizing photons at these
redshifts.  However, at $z < 3.1$, the data is consistent with a small
break strength ($B = 1 - 10$).  We shall consider this evolution with
redshift in more detail in section~4.6.

All of the \ion{N}{5}/\ion{C}{4} values shown in Fig.~9c correspond to
$z > 3.1$\ systems and the low values are then a necessary consequence
of large breaks across the He$^+$\ edge.  The presence of trace amounts
of \ion{N}{5} does imply that there must be some high energy ionizing
photons, or alternatively some high temperature gas in the systems.

\subsection{Strong Systems}

For the strongest systems ($N({\rm H~I}) \ge 5\ten{15}\cm2$) in Q1422+231 it
is possible to measure a larger suite of lines and also to obtain
measurements of, or at least upper limits to, ions that lie in the forest.  We
summarise in Table~4 our measurements of \ion{C}{2}, \ion{C}{3}, \ion{C}{4},
\ion{Si}{2}, \ion{Si}{3}, \ion{Si}{4}, and \ion{N}{5} in these systems.
\ion{C}{2} and \ion{Si}{2} are both low, as is \ion{N}{5}, as would be
expected from the strongly broken power law spectrum discussed in section~4.4.
\ion{Si}{3} is slightly weaker than \ion{Si}{4}, consistent with
$\log_{10}\Gamma \sim -1.5$, whereas \ion{C}{3}/\ion{C}{4} is less than 1.8,
which implies that $\log_{10}\Gamma > -2$.

\subsection{Si~IV/C~IV versus redshift}

We have searched in an augmented sample, which consists of the quasars
discussed above along with Q2000$-$330 ($z = 3.777$) and Q1623+268 ($z =
2.526$) for all systems having $N({\rm C~IV}) \ge 5\ten{12}\cm2$\ and for which
\ion{Si}{4} lies redward of the forest, as well as any system which is present
in \ion{Si}{4} and for which \ion{C}{4} lies within our wavelength coverage.
Our spectral coverage of the eight quasars provides a sample of 28 such
systems after excluding known Lyman limit systems.

We plot \ion{Si}{4}/\ion{C}{4} for these systems as a function of redshift in
Fig.~11, with upper limits shown as downward pointing arrows.  Below $z =
3.1$, 13 of 14 systems have \ion{Si}{4}/\ion{C}{4} less than 0.1 whereas 13 of
14 systems redward of $z = 3.1$\ have \ion{Si}{4}/\ion{C}{4} greater than 0.1.
A rank sum test rejects at the $4\ten{-5}$\ level the
possibility that the two samples are drawn from the same distribution
function.

As we have discussed in section~4.4 the transition most likely arises from the
change in ionizing spectrum from one at $z > 3.1$\ that has few photons above
the He$^+$\ edge to one with an unbroken power law at lower redshift.  The
transition is very abrupt --- a factor of at least ten in the break
strength in a redshift interval of less than 0.1; this favors a crossing of
the He$^+$\ ionization boundary in the IGM as the reason rather than the other
possibility, namely the changeover from a galaxy-dominated ionizing flux at $z
> 3.1$\ to a quasar-dominated flux at $z < 3.1$.

\ion{Si}{4}/\ion{C}{4} can be related to \ion{He}{2}/\ion{H}{1}, as we
illustrate in Fig.~12 for the power law-break models computed with CLOUDY.
At high redshift where \ion{Si}{4}/\ion{C}{4} is large,
\ion{He}{2}/\ion{H}{1} lies between 250 and 2000, depending on $\Gamma$,
whereas \ion{He}{2}/\ion{H}{1} should be much lower --- around 20 to 40 ---
at lower redshift ($z < 3.1$).

Kim et al.\ (1996) have shown recently that the number density evolution of
high and low column density clouds is quite distinct over this redshift range:
unlike high column density clouds, which show rapid evolution, the number
density of $N({\rm H~I}) < 3\ten{14}\cm2$\ clouds is invariant and described
by a $\beta \approx -1.5$\ power law.  It is these low column density clouds
that provide the dominant contribution of the forest to the He~II opacity.  If
we assume that the same metagalactic flux ionizes both low and high column
density clouds, then there will be an abrupt drop in the He~II opacity at $z
\sim 3.1$, with relatively constant values above and below this redshift.  The
drop in the opacity varies as (\ion{He}{2}/\ion{H}{1})$^{0.5}$\ for a $\beta =
-1.5$\ power law (Miralda-Escud\'e 1993) and so will lie between 2.5 and 10,
depending on typical values of $\Gamma$\ in the clouds.  This could explain
the discrepancy between the value of $\tau_{\rm He~II} >> 1.7$\ measured at $z
\le 3.286$\ (Jakobsen et al.\ 1994) and the much lower $\tau_{\rm He~II} = 1$\
found by Davidsen et al.\ (1996) at $z \le 2.72$.

\section{Proximate Systems}

The interpretation of the relative deficiency of $\lal$\ clouds close to a
quasar's redshift (the proximity effect) as being caused by the ionization of
the quasar itself (Bajtlik, Duncan \& Ostriker 1988, Lu, Wolfe \& Turnshek
1991) predicts that clouds within a few thousand $\kms$\ of the quasar should
have significantly higher ionization parameters than clouds in the general
field. This effect should be more pronounced for high ion ratios at $z > 3.1$\
as we have argued that the metagalactic ionizing flux at hWorkshopigh redshift is
severely deficient in high energy photons.

We have searched for this effect in our values of \ion{C}{4}/\ion{H}{1} and
\ion{Si}{4}/\ion{C}{4}, as is shown in Fig.~13 where we plot all clouds with
$N({\rm C~IV}) > 10^{12}\cm2$.  The cloud closest to Q1422+231 is indeed
highly anomalous.  This system in Q1422+231 shows
unsaturated $\lal$, extremely strong \ion{C}{4} and \ion{N}{5} but very weak
\ion{Si}{4} and \ion{C}{2}.  We show the spectra of $\lal$, \ion{C}{4},
\ion{Si}{4} and \ion{N}{5} in Fig.~14; the absorption consists of two roughly
equal components separated by $53\kms$\ whose properties are summarized in
Table~5.

The very low values of \ion{Si}{4}/\ion{C}{4} ($< 0.007$\ and 0.013
respectively) require a power-law ionization with a relatively small break at
the He$^+$\ edge; this of course is what is expected if the cloud is directly
exposed to the quasar's ionizing flux.  For a $\nu = -1.5$\ power law,
$\log_{10}\Gamma = -1.75$\ and $-1.5$\ produce good fits to the ionization
ratios in the two components.  However, with these values of the ionization
parameter, the observed \ion{C}{4}/\ion{H}{1} of 0.35 and 0.5 require an
approximately solar metallicity in the cloud, suggesting that it might be
associated with the quasar rather than lying in the general IGM.

The simplicity of the component structure in this system also allows us to test
the temperature prediction of the photoionization models.  Predicted
$b$-values versus atomic weight are shown in Fig.~15 with temperature
computed both with low metallicity (dotted line) and with solar metallicity.
In order to fit the observed trend in $b$, we must introduce a substantial
turbulent broadening equivalent to $b = 14\kms$\ in each component.

\section{The C~IV Correlation Function}

Fernandez-de Soto et al.\ (1996) have recently analyzed the \ion{C}{4} line
systems given in Paper~I and find a strong correlation signal in the
\ion{C}{4} absorption lines at small velocity separations.  The present
Q1422+231 data set provides a somewhat better determination of this. In
Figure~16 we show the two-point correlation function (TPCF) of \ion{C}{4}
absorption lines with $N({\rm C~IV}) > 10^{12}\cm2$\ and $2.66 < z < 3.62$.
The shape and strength of the correlation function are relatively insensitive
to the column density cut.  The correlation function is slightly weaker than
Fernandez-de Soto et al.\ find but it is in broad agreement with the results
of Womble, Sargent \& Lyons (1996) who analyzed  independent data on Q1422+231.

There is a very weak anticorrelation at $\Delta v > 300\kms$\ so it is not
possible to measure a large scale correlation signal with this data, and the
observed correlation at $\Delta v < 300\kms$\ is subject to the usual
uncertainties introduced by the internal velocity dispersions, effective
sizes, and geometries of the objects.  Moreover, it is clear from the
discussion of section~4.3 that \ion{C}{4} has a more extended velocity
distribution in a given line system than \ion{H}{1}, and so the $\lal$\
correlation function will have a weaker amplitude than that of \ion{C}{4}.

We fitted the TPCF with the functional form of Heisler, Hogan \& White (1989) 
\begin{equation}
\xi(\nu) = {{1} \over {(2\pi)^{1/2}}}\ \int^{\infty}_{r_{cl}}
{{H\,dr} \over {\sigma}} \, \left ( {{r} \over {r_{corr}}} \right ) ^{-\gamma}
\, \left [ \exp - {{(Hr - v)^2} \over {2\sigma^2}} 
+ \exp - {{(Hr + v)^2} \over {2\sigma^2}} \right ]
\end{equation}
\noindent
where $H(z)$\ is the Hubble constant at redshift $z$, $\sigma$\ is the
velocity dispersion of the clouds, $r_{cl}$\ is the cloud dimension and
$r_{corr}$\ is the correlation length.  The distances are in physical (proper)
units.  For $\qno = 0.5$, the best fit, shown as the dashed line in Fig.~15,
has $r_{corr} \approx 270 h^{-1}~{\rm kpc}$, $\sigma = 100\kms$, and $r_{cl} =
40 h^{-1}~{\rm kpc}$, where $h = (\hno /100\kms)$.  For comparison, Christiani
et al.\ (1995) found $r_{corr} = 280 h^{-1}~{\rm kpc}$, $\sigma = 150\kms$,
and $r_{cl} = 110 h^{-1}~{\rm kpc}$\ for the $\lal$\ correlation.  If we {\it
assume\/} that $\xi$\ corresponds to the galaxy correlation function and
evolves as 
\begin{equation}
\xi = (1 + z)^{-3-\epsilon} \ \left ( {{r} \over {r_0}} \right ) ^{-1.8}
\end{equation}
\noindent
where $r_0 = 5.5 h^{-1}~{\rm Mpc}$\ is the current galaxy correlation length,
then $\epsilon \sim 1$, which is slightly weaker than the $\epsilon = 2.4$\
found by Fernandez-de Soto et al.\ but still denotes a relatively fast
evolution in the galaxy correlation function, such as might be found in an
$\Omega = 1$\ universe.

\section{Conclusion}

We can summarise the conclusions fairly simply.  Lyman forest clouds with
$N({\rm H~I}) > 10^{15}\cm2$\ are ubiquitously metal-enriched, typically
having carbon abundances $10^{-2}$\ of solar and Si/C of about three times
solar, values very similar to those of Galactic halo stars.  At $N({\rm
H~I})$\ below $10^{15}\cm2$, curent sensitivity limits for $N({\rm C~IV})$\
are inadequate to determine whether or not all clouds have \ion{C}{4} if the
distribution function of \ion{C}{4}/\ion{H}{1} is similar to
that at higher column density.  The data are perfectly consistent with the
clouds having the same distribution function but could permit as many as half
of the clouds with $3\ten{14} < N({\rm H~I}) < 10^{15}\cm2$\ to be chemically
pristine. 

We have investigated the ionization in the $N({\rm H~I}) > 5\ten{14}\cm2$\
clouds and shown that they have relatively uniform properties with a spread of
at most an order of magnitude in their metallicities and ionization
parameters, the latter lying between $\Gamma = 10^{-2.4}$\ and $\Gamma =
10^{-1.5}$.  The one exception is the single cloud lying closer 
than $1000\kms$\ to a quasar, in which the metallicity is near solar.  The
clouds are structured in velocity, and the highest velocity components are
more highly ionized, which suggests that we are seeing layered structure in an
individual system such as might be formed in a pancake collapse.  This
internal structure may be responsible for the two point velocity correlation
which is seen at $\Delta v < 300\kms$.  Alternatively, if the velocity
correlation is interpreted in terms of the galaxy correlation function we find
a relatively rapid evolution of this quantity from $z = 3$\ to $z = 0$.

Finally, and perhaps most intriguingly, we find that there is a rapid
evolution in the value of \ion{Si}{4}/\ion{C}{4} at a redshift of about 3.1
which we interpret as being caused by a change in the spectrum of the
metagalactic ionizing flux, which must go from having a strong break at the
He$^+$\ edge at $z > 3.1$\ to having a weak break at $z < 3.1$.  This could be
a result of the transition from a galaxy-dominated to a quasar-dominated
ionizing flux, but, given the rapidity of the transition, it appears to be more
likely that we are seeing the He$^+$\ ionization bounary in the intergalactic
medium.  Irrespective of the origin of the effect, the evolution of the
ionizing spectrum implies that we will observe much weaker He$^+$\ edges in
quasar spectra at redshifts $z < 3.1$.

\appendix

\section{Structure of Individual Systems in Q1422+231}

Velocites are measured relative to the strongest \ion{H}{1} component.

{\it 3.264:}\quad \ion{C}{4} consists of one single broad line with
$N(\rm C~IV) = 4\ten{12}\cm2$\ and a $b$-value of $19\kms$.  It
corresponds to a hydrogen line with $N(\rm H~I) = 1.7\ten{15}\cm2$\ and
$b = 28\kms$.  The separation of the centroids of \ion{C}{4} and
\ion{H}{1} is $+3\kms$.

{\it 3.410:}\quad \ion{C}{4} consists of a stronger wide component with
$b = 22\kms$\ and $N(\rm C~IV) = 8\ten{12}\cm2$\ centered at $+1\kms$\
from the strongest \ion{H}{1} with $N(\rm H~I) = 1.8\ten{15}\cm2$\ and
$b = 26\kms$.  There is a weaker wing with $b = 14\kms$\ and $N(\rm
C~IV) = 2\ten{18}\cm2$\ at $-47\kms$\ from the main hydrogen component,
which appears to corresond to a blueward wing in the \ion{H}{1} at
$-43\kms$\ from the main component.  $N(\rm H~I) = 1.8\ten{14}\cm2$\ and
$b = 21\kms$\ in this wing.

{\it 3.499:}\quad There is a wide \ion{C}{4} line with $b = 19\kms$\ and
$N(\rm C~IV) = 2\ten{12}\cm2$\ at $+7\kms$\ from the \ion{H}{1} with $b =
26\kms$\ and $N(\rm H~I) = 2.7\ten{15}\cm2$.  The \ion{C}{4} column
density is somewhat insecure because of neighboring lines in the region of
\ion{C}{4} $\lambda 1548\ang$.

{\it 3.513:}\quad The \ion{C}{4} consists of two narrow components with
$b = 7\kms$\ and $N(\rm C~IV) = 7\ten{12}\cm2$\ and $b = 12\kms$, $N(\rm
C~IV) = 3\ten{12}\cm2$\ at $-5\kms$\ and $+20\kms$\ respectively from
the hydrogen line with $N(\rm H~I) = 1.8\ten{15}\cm2$\ and $b = 24\kms$.
Their weighted mean lies $+3\kms$\ redward of the neutral hydrogen.

{\it 3.536:}\quad This high column density system is extremely complex.
There are 8 main \ion{C}{4} clouds: ($-329\kms$, $b = 17\kms$, $N(\rm
C~IV) = 6\ten{12}\cm2$), ($-264\kms$, $b = 22\kms$, $N(\rm C~IV) =
4.4\ten{13}\cm2$), ($-172\kms$, $b = 10\kms$, $N(\rm C~IV) =
8\ten{12}\cm2$), ($-107\kms$, $b = 15\kms$, $N(\rm C~IV) =
10^{13}\cm2$), ($-86\kms$, $b = 11\kms$, $N(\rm C~IV) =
4.3\ten{13}\cm2$), ($-43\kms$, $b = 13\kms$, $N(\rm C~IV) =
4.6\ten{13}\cm2$), ($2\kms$, $b = 21\kms$, $N(\rm C~IV) =
6\ten{12}\cm2$), ($92\kms$, $b = 4\kms$, $N(\rm C~IV) = 10^{12}\cm2$) as
well as a further $3\ten{13}\cm2$\ distributed more uniformly.  The
strongest neutral hydrogen component has $N(\rm H~I) = 1.2\ten{16}\cm2$\
and $b = 22\kms$, but there are further strong components at negative
velocities: ($-263\kms$, $b = 34\kms$, $N(\rm C~IV) = 6.4\ten{15}\cm2$),
($-232\kms$, $b = 18\kms$, $N(\rm C~IV) = 2.0\ten{15}\cm2$),
($-177\kms$, $b = 22\kms$, $N(\rm C~IV) = 6.2\ten{15}\cm2$),
($-106\kms$, $b = 46\kms$, $N(\rm C~IV) = 3.4\ten{15}\cm2$).  The four
systems that have both \ion{C}{4} and \ion{H}{1} identified from
profile fitting ($-263\kms$, $-177\kms$, $-106\kms$, and $0\kms$) have
\ion{C}{4}/\ion{H}{1} = $6.9\ten{-3}$, $1.3\ten{-3}$, $3\ten{-3}$, and
$5\ten{-4}$ respectively.  Neutral hydrogen is quite weak in the extreme velocity
components:  $N(\rm H~I) = 7\ten{13}\cm2$\ at $-329\kms$\ for
\ion{C}{4}/\ion{H}{1} = $9\ten{-2}$, and $N(\rm H~I) = 2\ten{13}\cm2$\
for \ion{C}{4} = $5\ten{-2}$\ at $+92\kms$.  The two intermediate
components at $-86\kms$\ and $-43\kms$\ are more uncertain but consistent
with as much as $N(\rm H~I) = 4\ten{14}\cm2$\ in each.

{\it 3.565:}\quad  This system consists of a single \ion{C}{4} line with
$b = 13\kms$\ and $N(\rm C~IV) = 1.4\ten{12}\cm2$\ separated by
$+4\kms$\ from the $N(\rm H~I) = 2.5\ten{15}\cm2$\ and $b = 21\kms$\
neutral hydrogen line.

{\it 3.586:}\quad The main \ion{C}{4} component with $b = 23\kms$\ and
$N(\rm C~IV) = 1.6\ten{13}\cm2$\ lies at $+3\kms$\ from the neutral
hydrogen line with $b = 28\kms$\ and $N(\rm H~I) = 4.2\ten{15}\cm2$.  In
addition there is a probable second component with $b = 21\kms$\ and
$N(\rm C~IV) = 4\ten{12}\cm2$\ at $+144\kms$\ from this position.  (The
agreement between the doublets is not as precise as usual, and there is
a possibility that this component is spurious.)  There is only a weak
$\lal$\ feature near this velocity, with $b = 22\kms$\ and $N(\rm H~I) =
3.3\ten{13}\cm2$.  The \ion{C}{4} line lies at $+3\kms$\ from this line.

\smallskip
\acknowledgments
We are grateful to T. Bida, R. Campbell, P.  Gillingham, J. Aycock, T.
Chelminiak, B. Shaeffer and W. Wack for their extensive help in obtaining the
observations and to E.~Hu and K.~Roth who carried out parts of the observing.
We would like to thank P. Jakobsen and M. Shull for comments on an earlier
draft of this paper and Michael Rauch and Ray Weymann for helpful
discussions.

\newpage

\begin{figure}
\caption{The spectrum of Q1422+231.  The bottom panel shows a complete 
spectrum, while the upper is a $100~{\rm\AA}$\ region redward of the quasar's
$\lal$\ emission.  The data have been two-point smoothed.  This spectrum was
obtained on 1995 July~2,3 and August~5 using three settings of the HIRES
echelle spectrograph on the Keck telescope; the
total exposure was 8.33 hours.  The data were reduced as described in Paper~I
and Hu et al.\ 1995.  The spectrum has a resolution, R = 36,000,
and provides complete wavelength coverage from $3900~{\rm\AA}$\ to
$7200~{\rm\AA}$.  At the C~IV region shown, the S/N is measured as 160 per
resolution element.  31 C~IV systems containing 66 velocity components can be
identified in the C~IV region longward of the Lyman forest ($z = 2.66 -
3.63$).}
\end{figure}

\begin{figure}
\caption{Atlas of 62 ${\rm L}{\alpha}$-saturated clouds in Q1422+231 for 
$z > 2.65$.  In each case we show the ${\rm L}{\alpha}$\ line
(bottom), then the ${\rm L}{\beta}$\ line (for systems with $z >
2.92$), and then the C~IV $\lambda 1548~{\rm\AA}$\ and C~IV$\lambda
1550~{\rm\AA}$\ lines (top).  The observed air redshift is shown in the left
corner.}
\end{figure}

\begin{figure}
\caption{A $100\ang$\ portion of the spectrum of Q1422+231 (bottom) with the
corresponding ${\rm L}\beta$\ (middle) and C~IV $\lambda 1548\ang$\ (top)
shifted in wavelength so they lie above ${\rm L}\alpha$.  Dotted vertical
lines indicate clouds that are saturated in both ${\rm L}\beta$\ and ${\rm
L}\alpha$.  Short dotted lines show the position of C~IV $\lambda 1550\ang$\
corresponding to these clouds.}
\end{figure}

\begin{figure}
\caption{(a)\quad Column densities of C~IV corresponding to all Lyman forest systems
with saturated ${\rm L}\beta$\ in numerous quasar lines of
sight.  These are --- Q0014+813: $z = 2.84 - 3.30$; Q0256$-$000: $z = 2.79 -
2.86$; Q0302$-$003: $z = 2.74 - 2.86$; Q0636+680: $z = 2.64 - 2.74$;
Q0956+122: $z = 2.75 - 2.92$; Q1159+123: $z = 2.95 - 3.05$; Q1206+119: $z =
2.60 - 2.85$; Q1422+231: $z = 2.92 - 3.60$; Q2126$-$158: $z = 2.73 - 3.00$.
Detected C~IV systems are shown at the measured column density but all
undetected systems are shown at a nominal value of $10^{12}~{\rm
cm}^{-2}$. Throughout the figures, absorption lines toward Q1422+231 are shown
as diamonds, those toward Q0014+813 as triangles, and the remainder as
squares.   \quad (b)\quad As in (a) for Si~IV; undetected systems are shown
at $5\ten{11}\cm2$. \quad (c)\quad As in (a) for N~V; undetected systems are
shown at $10^{12}\cm2$. \quad (d)\quad As in (a) for C~II; undetected systems
are shown at $10^{12}\cm2$.}
\end{figure}

\begin{figure}
\caption{$N({\rm C~IV})$\ versus redshift for $\lal$\ saturated clouds in
Q1422+231.  The solid line shows the adopted detection limit of
$5\times 10^{11}\cm2$. }
\end{figure}

\begin{figure}
\caption{The detected fraction of clouds versus the parameter $x \equiv N({\rm
C~IV})_{det}/N({\rm H~I})_{select}$, where $N({\rm C~IV})_{det}$\ is the
detection limit for C~IV and  $N({\rm H~I})_{select}$\ is the selection limit
in $N({\rm H~I})$.  The filled squares and diamonds show the detection
fraction for the $\lal$-selected sample [$N({\rm H~I})_{select} \approx
3\ten{14}\cm2$] with $\pm 1\sigma$\ error bars, whereas the solid line shows
the result for the $\lbet$-selected sample [$N({\rm H~I})_{select} \approx
1.6 \times 10^{15}\cm2$].}
\end{figure}

\begin{figure}
\caption{C~IV versus H~I column density for all systems with
$N({\rm H~I}) > 5 \times 10^{14}~{\rm cm}^{-2}$\ at $3.135 < z < 3.60$\
toward Q1422+231(diamonds) and for $N({\rm H~I}) \ge 1.5
\times 10^{15}~{\rm cm}^{-2}$\ at $z > 2.95$\ toward Q0014+813 (triangles).
Also shown (open squares) are detected C~IV/H~I values in all PLLSs 
toward the eight quasars (Songaila 1996).  The dashed line shows the
typical $2~\sigma$\ detection limit for C~IV in Q1422+231.  The solid lines
show model calculations (Bergeron \& Stasinska 1986) of C~IV/H~I for $\Gamma =
10^{-2.7}$\ (lower), $\Gamma = 10^{-1.7}$\ (upper) and the dotted line the
model for $\Gamma = 10^{-0.7}$, and a metallicity of $10^{-2}$\ solar, and
illustrate that the difference between the PLLS and lower column density
systems is not solely a radiative transfer effect but must arise from a higher
ionization parameter or higher metallicity in the weaker clouds.}
\end{figure}

\begin{figure}
\caption{(a)\quad C~IV/H~I versus H~I for the components described in
Appendix~A.\quad\quad (b)\quad C~IV/H~I versus velocity separation from the
strongest H~I component in each absorption line system.}
\end{figure}

\begin{figure}
\caption{Si~IV/C~IV, C~II/C~IV and N~V/C~IV as a function of C~IV for all
${\rm L}{\beta}$\ saturated systems where the lines lie redward of the forest
and within the wavelength coverage (filled squares).  
The solid lines are $2\sigma$\ detection limits for Q1422+231.  Upper limits
for the other quasar lines of sight are shown with downward pointing arrows.
We have also shown values for the PLLSs (open squares).
}
\end{figure}

\begin{figure}
\caption{Comparison of the observed values of Si~IV/C~IV versus C~II/C~IV with
model predictions of the CLOUDY code.  The models are computed for a
$\nu^{-1.5}$\ power law with various breaks across the He$^+$\ edge (dashed
line --- no break; dotted line --- factor of 2; dash-dot line --- factor of
10; long dash line --- factor of 100; solid line --- factor of 1000) and for
two values of Si/C (left panel --- solar; right panel --- low metallicity,
Si/C = 3 times solar).}
\end{figure}

\begin{figure}
\caption{Si~IV/C~IV versus redshift for all clouds with $N({\rm
C~IV}) \ge 5\times 10^{12}~{\rm cm}^{-2}$\ that are not known Lyman limit
systems and for which Si~IV lies longward of the Lyman alpha forest.
(Downward pointing arrows are $1~\sigma$\ upper limits.)  The filled diamonds
are values for Q1422+231 and the filled triangles for Q0014+813.  The squares
are from the sample of quasars of Table~1, augmented at the lowest redshift by
observations of Q1623+269 (open squares) and at the higher redshifts with
Q2000$-$330, which are shown by the open triangles.  The tick marks show the
redshifts below which He$^+$\ edge breaks have been measured (Jakobsen et al.\
1994, Tytler et al.\ 1995, Davidsen et al.\ 1996).  The optical depth measured
by Davidsen et al.\ (1996) at $z = 2.743$\ is much lower ($\tau = 1$) than that
measured by Jakobsen et al.\ (1994) at $z = 3.286$\ ($\tau >> 1.7$); these
redshifts roughly bracket the epoch at which Si~IV/C~IV values drop abruptly.}
\end{figure}

\begin{figure}
\caption{He~II/He~I versus Si~IV/C~IV for the CLOUDY models of Fig.~10.  We
assume here the low metallicity ratio for Si/C.}
\end{figure}

\begin{figure}
\caption{(a) C~IV/H~I and (b) Si~IV/C~IV versus separation velocity from the
quasar for all clouds with $N({\rm C~IV}) > 10^{12}~{\rm cm}^{-2}$.  Clouds
toward Q1422+231 are shown as diamonds, and clouds toward Q0014+813 are shown
as triangles.}
\end{figure}

\begin{figure}
\caption{Line profiles for the proximate system at $z = 3.623$\ in the
spectrum of Q1422+231.  The bottom spectrum is Ly$\alpha$\ with above (in
order) C~IV ($\lambda 1548~\rm\AA$), C~IV ($\lambda 1550~\rm\AA$), Si~IV ($\lambda
1393~\rm\AA$), Si~IV ($\lambda 1402~\rm\AA$), N~V ($\lambda 1238~\rm\AA$),
N~V ($\lambda 1242~\rm\AA$).}
\end{figure}

\begin{figure}
\caption{$b$-value versus atomic weight for the two velocity components of
the system of Fig.~13 (Ly$\alpha$, C~IV and N~V) are shown as squares.  The
dotted and dashed lines show model predictions for the thermal broadening at
temperatures predicted by the CLOUDY models for low metallicity (dotted) and
solar metallicity (dashed).  The solid line includes a further turbulent
component with $b = 14~{\rm km\ s}^{-1}$.}
\end{figure}

\begin{figure}
\caption{The two-point velocity correlation function of the 52 C~IV
lines with $N({\rm C~IV}) \ge 3\times 10^{12}~{\rm cm}^{-2}$\ toward
Q1422+231, excluding the partial Lyman limit system.  The correlation function
is shown in $50~{\rm km\ s}^{-1}$\ bins with $1~\sigma$\ error bars based on
the number of pairs in each bin.}
\end{figure}

\newpage

\centerline{TABLES}\bigskip\noindent
TABLE~1 --- C~IV Lines in Q1422+231 for $N({\rm H~I}) > 5\ten{14}\cm2$\
Systems\par\noindent 
TABLE~2 --- C~IV Lines in Q0014+813 for $N({\rm H~I}) > 10^{15}\cm2$\
Systems\par\noindent 
TABLE~3 --- Partial Lyman Limit Systems\par\noindent
TABLE~4 --- Strong Systems in Q1422+231\par\noindent
TABLE~5 --- Proximate System

\end{document}